\documentclass[letter]{aa}
\usepackage{gensymb}

\usepackage{geometry}
\usepackage{soul}
\usepackage{graphicx}
\usepackage{color}
\usepackage{wrapfig}

\usepackage{txfonts}

\usepackage{bbold}
\usepackage{caption}
\usepackage{subcaption}

\usepackage{here}
\usepackage{tikz} 
\usepackage[fleqn]{amsmath}
\usepackage{mathrsfs}

\usepackage[colorlinks=True,pdfborder={0 0 0}, linkcolor=blue,citecolor=blue,breaklinks=true]{hyperref}

\usepackage{mwe}

\begin{document}

  \renewcommand{\vec}[1]{\mathbf{#1}}

   \title{Evading the dust fragmentation barrier with the streaming instability in protoplanetary disks} 
  
   \author{ V. Vallucci-Goy
          \inst{1} \and
          U. Lebreuilly 
          \inst{1} \and
          M.-M. Mac Low 
          \inst{2} \and
          P. Hennebelle 
          \inst{1} }

   \institute{Université Paris-Saclay, Université Paris Cité, CEA, CNRS, AIM, 91191, Gif-sur-Yvette, France. \and
   Department of Astrophysics, American Museum of Natural History, New York, NY 10024, USA. \\
              \email{valentin.vallucci@gmail.com}
             }

   \date{}

 
  \abstract
   {The streaming instability (SI) is a leading candidate for reaching solid densities sufficient to trigger the gravitational collapse needed for the formation of planetesimals. However, dust growth barriers appear to impede the ability to assemble sufficiently large dust particles to trigger strong clumping, providing a serious impediment to planetesimal formation.}
   {We aim to address the possibility to enhance dust clumping with dust growth in SI-produced structures, and to estimate the impact of the shift of the dust fragmentation threshold in regions where the SI has enhanced the dust density.}
   {We perform two-dimensional numerical simulations of the SI with a monodisperse description of dust growth, accounting for the impact of mass loading of the dust on the sound speed of the gas and dust mixture when computing dust collisional velocities.}
   {Dust mass loading reduces collision velocities in high density regions, allowing dust particles to survive to larger sizes before shattering. In turn, dust clumping is boosted as particles grow in size, as long as they remain sufficiently coupled to the gas. }
   {This two-way synergy between dust growth and clumping, which depends on the initial dust-to-gas ratio and dust elastic properties, allows denser dust clumps to form  and thus facilitates the onset of planetesimal formation.}

   \keywords{Hydrodynamics; Magnetohydrodynamics (MHD); Turbulence;  Protoplanetary disks; Planets and satellites formation.}

    \authorrunning{Vallucci-Goy et al.}
    \titlerunning{Evading the fragmentation barrier with the SI}

    \maketitle

%
\section{Introduction}

In protoplanetary disks, the pathway to planetesimals remains unclear, although the first generation is believed to form quite early in the disk lifetime \citep{Testi2014} in a gas-rich environment via local gravitational instability (GI) \citep{Goldreich1973} of solid material. The leading candidate to produce dust density fluctuations strong enough for the onset of GI, is the streaming instability (SI), which is fuelled by relative motions of solid particles and gas molecules that communicate via aerodynamic drag. This runaway convergence of radial drifts was first studied analytically in the linear regime by \cite{Youdin2005}, and then numerically for the first time by \cite{Johansen2007} in the non-linear regime, showing that filaments of high dust density form in a background of quasi-incompressible gas.
However, the growth of pebbles to high enough Stokes numbers $\mathrm{St}=t_{\rm s}  \Omega$, the product of the stopping time and the orbital frequency, to trigger the SI is hindered by the so-called bouncing, fragmentation, and radial drift barriers \citep{Birnstiel2024}. 

Arguments have been made that the snow line is a privileged place for planetesimal formation \citep{Drazkowska2014,Schoonenberg2017,Drazkowska2017}. The role of pressure gradients was explored in \cite{Bai&Stone2010,Abod2019} and that of external turbulence in \cite{Umurhan2020} and \cite{Lim2024}. \cite{Carrera2025b,Huhn2025} investigate how planetesimal formation responds to turbulence driven by accretion in the very first stage of disk evolution dominated by infall. Dust self-gravity was included in \cite{Johansen2007,Johansen2015,SimonArmitage2016,Simon2017,Schafer2017,Li2019} along with vertical stratification to self-consistently follow and describe the initial planetesimal mass function. The impact of magnetic effects was studied in \cite{Lin2022}. 

A broad parameter space of initial values of  St and surface dust-to-gas ratio $Z_\mathrm{crit}$ was extensively studied in the literature \citep{Carrera2015,Yang2017,LiYoudin2021,Carrera2022,Lim2024} in order to define the conditions needed for strong clumping, i.e.\ to exceed the Roche density $\rho_\mathrm{R}$ and trigger the gravitational instability locally. Taking into account a full size distribution of dust particles, it was shown that growth rates of the instability tend to be reduced compared to the single species case \citep{Paardekooper2020,Paardekooper2021}. However, this question is still under debate with no apparent convergence of the dust dynamics and SI behaviour with the number of bins considered in the dust size distribution model \citep{Schaffer2018,Paardekooper2021,Schaffer2021,Zhu&Yang2021,YangZhu2021,Rucska2023}. Thus, planetesimal formation seems more challenging than previously believed. 

Recently, velocity dispersions of solid particles were measured in simulations \citep{Schreiber2018} to estimate collisional velocities and dust coagulation rates \citep{Tominaga2023,Tominaga2025}, reaching the conclusion that collisional velocities driven by the coupling to turbulent vortices in the gas should be lower than those predicted by the widely used analytical sub-grid model of \cite{Ormel2007}. Through analytical arguments, the work of \cite{Carrera2025a} suggests a positive synergy between dust coagulation and clumping by the SI, accounting for the back reaction produced by the mass loading of dust particles and the resulting attenuation of turbulence. This allows relaxation of the conditions for strong clumping. 

Dust growth was included in simulations of SI  for the first time, along with vertical stratification, by \citet{Ho2024}. They show that dust clumping due to SI boosts dust growth, which in turn leads to SI modes associated with stronger density enhancements that exceed the Roche density more easily.

In this work, we perform high-resolution 2D simulations of the SI within a shearing box including monodisperse dust growth, accounting for sticking and fragmentation. We use the turbulent collisional velocities damped by dust mass loading to model dust coagulation and thus to investigate the possibility of a fragmentation barrier occurring at larger size than predicted by \citet{Ormel2007}, in turn producing stronger dust clumping.

\section{Methods}
\label{methods}
We use the grid-based {\ttfamily SHARK} code \citep{Lebreuilly2023SHARK} to perform 2D simulations of a mixture of unmagnetized isothermal gas and a single dust species treated as a fluid, without vertical stratification. We model a local patch of the protoplanetary disk centered on the midplane within the shearing box framework \citep{Goldreich1978}. The momentum equations for the gas and dust mixture are given in App.~\ref{appendix:momentum equations}. They include the hydrodynamic drag force between dust and gas $\vec{F}_\mathrm{drag} = (\rho_\mathrm{d}/{t_{\rm{s}}} ) (\vec{v}_\mathrm{d}-\vec{v})$, where $\rho_\mathrm{d}$ and $\vec{v}_\mathrm{d}$ are the dust density and velocity, while $\vec{v}$ is the gas velocity. The aspect ratio of the disk $\left(H/r\right) = 0.1$ is taken constant with radius. We work in code units and set $\Omega = r = \rho_\mathrm{g,0} = 1$, where $\rho_\mathrm{g,0}$ is the initial gas density.  This gives us a locally isothermal sound speed $c_\mathrm{s} = \left(H/r\right) v_\mathrm{k} = H \Omega = 0.1$, where $v_\mathrm{k}$ is the Keplerian velocity. The stopping time is taken in the \cite{Epstein1924} regime where $t_\mathrm{s} = (8/\pi)^{1/2} (\rho_\mathrm{gr}s_\mathrm{d})/(\rho_\mathrm{g}  c_s),$ where we take $\rho_\mathrm{gr} / \rho_\mathrm{g}  = 10^{11}$ to define the grain intrinsic density $\rho_\mathrm{gr}$ kept constant throughout the simulation (see App. \ref{sect:domain size}) , so  the  Stokes number St $
\propto s_\mathrm{d}$, the dust size.

\subsection{Dust growth}

We use a monodisperse approach to follow at low cost the evolution of a single dust species that follows the maximum of the dust size distribution. Assuming that the bulk of the mass is carried by the largest grain of the dust distribution, its evolution is given by \citep{Stepinski1996,Ormel2008,Estrada2008,Sato2016,Tsukamoto2017}
\begin{equation} \label{monodisperse equation}
    \frac{\mathrm{d}\mathrm{St}}{\mathrm{d}t} = \frac{\mathrm{St}}{3t_\mathrm{coag}},
\end{equation}
where coagulation is driven by collisions of equal-size particles. The corresponding timescale is
\begin{equation}
    t_\mathrm{coag} \equiv \frac{1}{K n_\mathrm{d}} = \frac{m_\mathrm{gr}}{({8}/{3 \pi})^{1/2} 4 \pi s_\mathrm{d}^2  \Delta v_\mathrm{turb} \epsilon \rho_\mathrm{g}},
    \label{eq:t_coag}
\end{equation}
where $K $ is the collision kernel, $n_\mathrm{d}$ is the  dust number density, $m_\mathrm{gr} \equiv (4/3) \pi s_\mathrm{d}^3 \rho_\mathrm{gr}$ is the mass of the dust grain, $\Delta v_\mathrm{turb}$ is the turbulent induced velocity, and $\epsilon = \rho_\mathrm{d}/\rho_\mathrm{g}$ is the dust-to-gas mass ratio.

When fragmentation is included, complete shattering of the dust grains occurs whenever the collision velocity 
\begin{equation} \Delta v_{\mathrm{turb}} > v_\mathrm{frag},
\label{frag condition}
\end{equation}
where the fragmentation threshold velocity is $v_\mathrm{frag}$.

\subsection{Collision velocity}
\label{sect:collision velocity}
We consider turbulence as the source of grain collisions \citep{Ormel2007}. Following \cite{Carrera2025a}, we use a modified sound speed $\tilde{c}_\mathrm{s}$ reduced by the mass loading of the gas by the dust particles. Assuming that turbulence is generated by a finite energy source that is partitioned between the gas and the dust,
\begin{equation}
\label{modified sound speed}
 \tilde{c}_\mathrm{s} = \frac{c_\mathrm{s}}{(1 + \epsilon/[1 + \mathrm{St}])^{1/2}}. 
\end{equation}
This mechanism is enhanced in regions of high dust concentration but saturates as St increases. The resulting relative velocity of each pair of grains in the intermediate regime
\begin{equation}
\label{modified Ormel eq}
    \Delta v_{\mathrm{turb}} \equiv \tilde{V}_\mathrm{g} (1.97 \mathrm{St})^{1/2} = \tilde{c}_\mathrm{s} (\alpha 1.97 \mathrm{St})^{1/2},
\end{equation}
where $\tilde{V}_\mathrm{g} \equiv \alpha^{1/2} \tilde{c}_\mathrm{s}$ is the modified average gas velocity for a \citet{Shakura76} turbulent viscosity parameter of $\alpha$ that we take in our models as $10^{-4}$, consistent with observations from ALMA \citep{Villenave2022} and values measured in SI simulations in \cite{Bai&Stone2010} and \cite{Schreiber2018}. We note that this environment-dependent parameter is important, as a value of $10^{-3}$, corresponding to more intense turbulence, would lead to collision velocities about three times larger. 

Using the fragmentation condition (Eqs.\ \ref{frag condition} and~\ref{modified Ormel eq}), we see that mass loading allows the fragmentation barrier $\mathrm{St_{frag}} = v_\mathrm{frag}^2/1.97{V_\mathrm{g}}$ to be pushed to higher values $\mathrm{St_{frag}} \rightarrow \tilde{\mathrm{St}}_\mathrm{frag} > \mathrm{St_{frag}}$
\begin{equation}
\tilde{\mathrm{St}}_\mathrm{frag} = \frac{v_\mathrm{frag}^2}{1.97\tilde{V_\mathrm{g}}} = \frac{v_\mathrm{frag}^2} {1.97 V_\mathrm{g}} \sqrt{1 + \frac{\epsilon}{1 + \tilde{\mathrm{St}}_\mathrm{frag}}}
\end{equation}
This is a second-order polynomial that yields a single physically relevant root given in \cite{Carrera2025a}.

\subsection{Parameters}
\begin{table}
\caption{Models simulated, with a $2048^2$ resolution.}
\begin{center}
\begin{tabular}{lrlr}
\hline Model & $v_\mathrm{frag}/c_\mathrm{s}$ & $\epsilon_0$ & $L_x=L_z$ 
\\
 \hline
  $v_\mathrm{frag}0.007 $&$7\times 10^{-3}$&$0.3$&$0.94H$\\
  $v_\mathrm{frag}0.015 $&$1.5\times 10^{-2}$&$0.3$&$0.94H$\\
 $\epsilon0.1 $&$1.5\times 10^{-2}$&$0.1$&$0.94H$\\
  $\epsilon1 $&$1.5\times 10^{-2}$&$1$&$0.094H$\\

\hline

\end{tabular}
\tablefoot{For each parameter choice, three cases have been run: NGr, GrNML, and GrML.}
\end{center}
\label{table1}
\end{table}
 We explore three different values of initial dust-to-gas ratio $\epsilon_0 = \left[0.1,0.3,1  \right]$ following \citet{Johansen2007}, which translate into the following range of surface dust-to-gas ratio \citep[][]{Lim2024} $Z = \Pi \epsilon_0/2 = \left[0.0025,0.0075,0.025 \right]$. Fragmentation velocities are chosen in the range $v_\mathrm{frag} = \left[7 \ \times 10^{-3}, 1.5 \ \times 10^{-2} \right] c_\mathrm{s}$ (justified in App.\ \ref{sect:domain size}). Regarding the domain size, we rely on \cite{Zhu&Yang2021} to estimate the wavelength of the fastest growing modes (more details in App.\ \ref{sect:domain size}). Information about the different models is gathered in Table \ref{table1}. 
 
We focus on the impact of dust growth and mass loading on the saturation state of the SI by comparing simulations without growth (NGr), with growth but no mass loading (GrNML) and with mass loading (GrML). Starting with an initial $\mathrm{St_{ini} = 0.1}$, dust growth is turned on when the saturation state of the SI is reached. Without mass loading, the dust distribution is expected to halt at a maximum $\mathrm{St_{frag}}$ determined by the choice of $v_\mathrm{frag}$. The correspondence between the two is given in Fig.\ \ref{fig:turb velocity vs St}. If dust clumping is strong enough, the fragmentation barrier is expected to be pushed to a higher Stokes number (i.e.\ higher dust size) when mass loading is accounted for. 

\section{Results}
\label{Results}

\begin{figure}[]
    \includegraphics[width=0.5\textwidth]{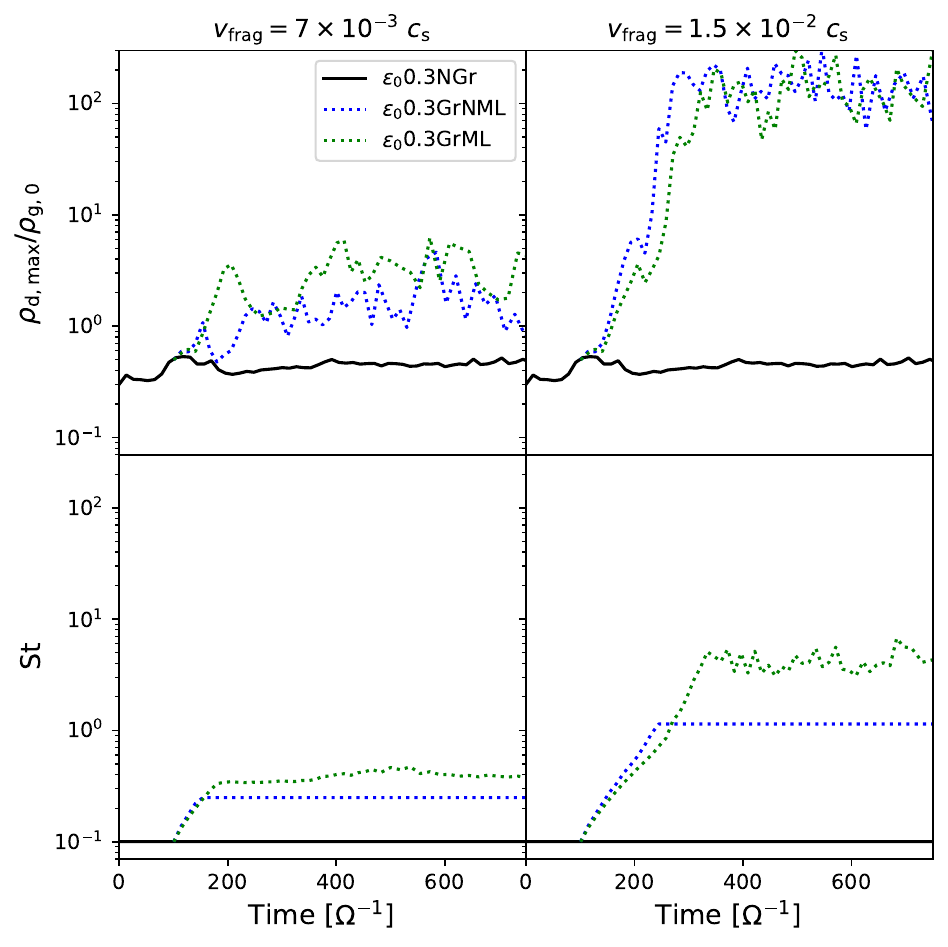}
    \caption{Evolution of the maximum dust-to-gas ratio {\em (top row)} and maximum St {\em (bottom row)} with time for the different dust growth scenarios NGr, GrNML, and GrML. The 
    velocity fragmentation threshold 
    is given above each column. In all cases, the initial dust-to-gas ratio $\epsilon_0 = 0.3$.}
    
\label{fig: rhod and St for eps0.3}
\end{figure}

\begin{figure*}[hbt]
    \centering
    \includegraphics[width=0.8\textwidth]{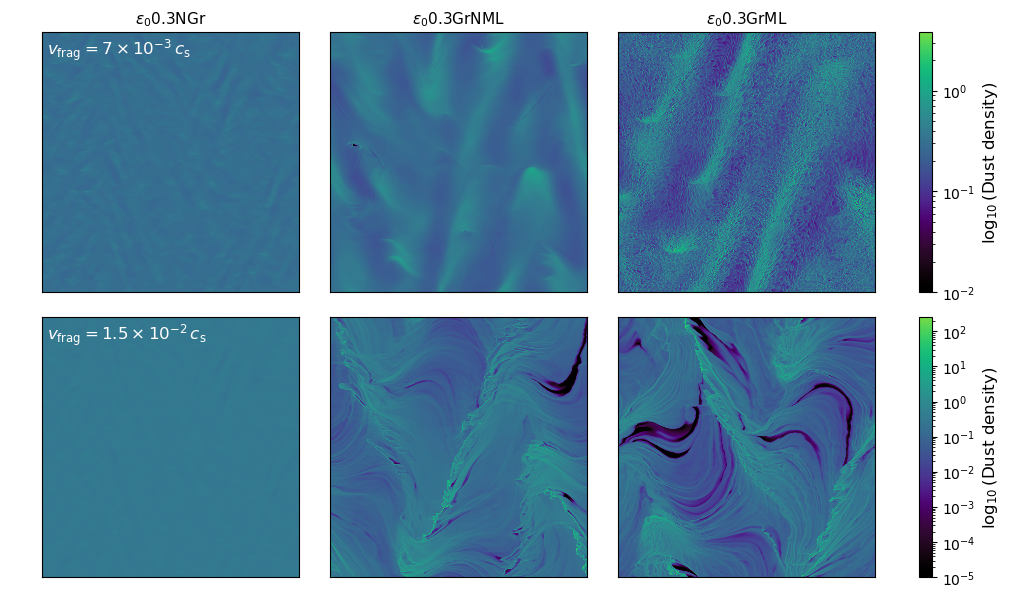}
    \caption{Dust density maps taken at saturation. Each column corresponds to the given dust growth scenario, while each row has the given 
    fragmentation velocity.}
    
\label{fig:density map}
\end{figure*}

We present in Fig.\ \ref{fig: rhod and St for eps0.3} the maximum dust-to-gas ratio as a function of time for different dust growth scenarios and two different fragmentation thresholds, with $\epsilon_0 = 0.3$. Figure \ref{fig:density map} displays dust density maps in the saturation state of the SI. 

First of all, looking at the run without dust evolution NGr, we see that only modest density fluctuations occur because both $\epsilon_0$ and St are below unity, corresponding to model AA in \citet{Johansen2007}. Including dust evolution in runs GrNML and GrML induces a striking rise in the maximum dust-to-gas ratio, owing to the increase in dust size, and thus St, triggering stronger SI. This increase is orders of magnitude greater for a higher fragmentation velocity, in agreement with the results of \cite{Ho2024}. The impact of mass loading of dust particles differs depending on the chosen fragmentation velocity.

\subsection{Tightly coupled regime St$_\mathrm{frag} < 1$}
Considering the lower fragmentation velocity $v_\mathrm{frag} = 7 \times 10^{-3} c_\mathrm{s}$, the Stokes number of the dust grows to a fixed value $\mathrm{St}_\mathrm{frag}\simeq 0.25 < 1$ for the GrNML run, as seen in Fig.\ \ref{fig: rhod and St for eps0.3} (and Fig.\ \ref{fig:turb velocity vs St}). Accounting for mass loading, the GrML run reaches a higher $\tilde{\mathrm{St}}_\mathrm{frag}$ that can be estimated by using Eq. (8) in \cite{Carrera2025a} and taking the $\epsilon$ value at the time where $\mathrm{St} \simeq \mathrm{St}_\mathrm{frag}$. Here, this corresponds to $\epsilon = \rho_\mathrm{d,max}/\rho_\mathrm{g}  \simeq 1$ (at $t\simeq 150 \ \Omega^{-1}$), which yields $\tilde{\mathrm{St}}_\mathrm{frag} \simeq 0.4$, a value close to that shown in Fig.\ \ref{fig: rhod and St for eps0.3}. In this tightly coupled regime with $\mathrm{St_{frag}} < 1$, increasing the maximum St of the dust allows triggering of more intense modes, which is a known behaviour of the SI \citep{Youdin2005}. Indeed, while the GrNML run reaches $\rho_\mathrm{d,max} / \rho_\mathrm{g}  \simeq 1$, this value is enhanced to $\sim 3$ in the GrML case, highlighting the synergy between mass loading and dust clumping.

\subsection{Loosely coupled regime $\mathrm{St_{frag}} \geq 1$} 
\label{subsec:loose} 
Considering now the higher fragmentation velocity $v_\mathrm{frag} = 1.5 \times 10^{-2} c_\mathrm{s}$, the Stokes number of the dust halts at  $\mathrm{St}_\mathrm{frag} \simeq 1$ for the GrNML run. Because of dust clumping, this value increases to $\tilde{\mathrm{St}}_\mathrm{frag} \simeq 4 > 1$ in the GrML run. While the SI is exacerbated when $\mathrm{St} \rightarrow 1$ due to a resonance effect \citep{Youdin2005,Hopkins2018a}, further decoupling of the dust actually leads to weaker density fluctuations. Indeed, although $\tilde{\mathrm{St}}_\mathrm{frag} \sim 4 \mathrm{St}_\mathrm{frag}$, the dust is now in a loosely coupled regime so its density is not further enhanced. Both runs GrNML and GrML lead to a maximum dust-to-gas ratio as exceeding 100.

\section{Discussion}
\label{discussion}

\subsection{Synergy between SI and dust growth}

We identified two regimes controlled by the chosen value of fragmentation velocity. Since dust tensile properties are not tightly constrained \citep[see][]{Gundlach2015,Kimura2015,Steinpilz2019,Gartner2017,Musiolik2021,Pillich2023,Morrissey2025} and dust surface composition varies across the radial extent of the disk, we cannot determine yet which specific coupling regime should be expected at each radius. However, it is clear that as long as fragmentation occurs for dust particles with $\mathrm{St}_\mathrm{frag} < 1$, regardless of the source of collisions, mass loading in dust-rich regions should significantly reduce impact velocities and thus shift the fragmentation barrier to higher sizes. In turn, this should cause filaments with higher dust density to form. Thus, there is a two-way synergy between dust clumping induced by the SI and dust growth, as analytically suggested by \cite{Carrera2025a}. However, particularly resilient dust grains that have $\mathrm{St}_\mathrm{frag} \geq 1$ decouple from the SI, so do not further boost clumping, removing the two-way synergy. 

\subsection{Roche density and bouncing barrier}
In the tightly coupled regime, mass loading enhances dust density fluctuations, reducing the gap between the peak density and that required for gravitational instability. However, the efficiency of this mechanism depends on the reservoir of solid particles. In Figs.\ \ref{fig: rhod and St for varying eps} and \ref{fig:1D pdf}, we vary the initial dust-to-gas ratio and show that a value of $\epsilon_0 = 1$ rapidly pushes the dust from $\mathrm{St}_\mathrm{frag} \simeq 1$ to $\tilde{\mathrm{St}}_\mathrm{frag} \simeq 10$ along with a maximum dust-to-gas ratio close to $\sim 400$. For an aspect ratio $\left(H/r\right) = 0.1$ at $r = 1 \ \mathrm{AU}$, and an MMSN density profile given in \cite{Hayashi1981}, the Roche density is $\rho_\mathrm{R} = 9\Omega_\mathrm{k}^2/\left(4\pi \mathcal{G}\right) \simeq 374 \rho$, a rather high value, but reachable. Note that with a different choice of  $\left(H/r\right) = 0.05$ at $r = 10 \ \mathrm{AU}$, we get $\rho_\mathrm{R} = 9\Omega_\mathrm{k}^2/\left(4\pi \mathcal{G}\right) \simeq 59 \rho$, which is a value already achievable without mass loading \citep{Ho2024}. Simulations with vertical gravity (to account for settling of large particles toward the disk midplane) and self-gravity are needed to assess the full potential of dust mass loading. 

In Fig.\ \ref{fig: rhod and St for bouncing barrier}, we use the exact same approach but crudely explore the bouncing barrier by setting a lower initial $\mathrm{St_{ini} = 10^{-2}}$ and a bouncing velocity threshold $v_\mathrm{bouncing} = 1.5 \times 10^{-3} \ c_\mathrm{s}$, ten times lower than the fiducial $v_\mathrm{frag}$, corresponding to less favorable conditions for the development of the SI. While the GrNML run for $\epsilon_0 = 0.3$ produces negligible dust clumping in this scenario, the GrML scenario stands out by showing some clumping, though more than an order of magnitude less than earlier runs. For $\epsilon_0 = 0.1$, no density enhancement is observed.
\subsection{Radial drift}
As $\mathrm{St} \rightarrow 1$, the drift timescale reaches a minimum and radial drift becomes a serious impediment to the formation of planetesimals as pebbles are lost to the central star \citep{Armitage2019,Birnstiel2024}. However, radial drift should be less efficient in dust clumps \citep{Bai&Stone2010}. \cite{Tominaga2023} suggested that the increase in $\epsilon$ would compensate that of St such that drift velocities should not increase as dust growth proceeds. In addition, a high $\epsilon_0 = 1$ induces  fast dust growth and the formation of decoupled particles with $\tilde{\mathrm{St}}_\mathrm{frag} \simeq 10 > 1$ due to mass loading (see Fig.\ \ref{fig: rhod and St for varying eps}). In this case, drift velocities should decrease and radial drift should be less dramatic. Specific studies are needed to determine if dust traps are still needed to bypass this barrier \citep{Birnstiel2013}.

\section{Conclusion}
\label{Section conclusion}

Through high-resolution simulations including a monodisperse treatment of dust growth, we showed that dust clumping is boosted by dust coagulation as the Stokes number increases towards unity, suggesting a positive feedback loop. In addition, we explored the impact of mass loading of dust particles onto the gas, which is significant in high-dust-density regions. This mechanism reduces impact velocities and thus allows dust grains to survive to higher sizes, raising the fragmentation threshold. Depending on the velocity threshold for fragmentation, two regimes stand out:

\begin{itemize}
    \item If the fragmentation threshold is such that the Stokes number of fragmenting grains $\mathrm{St_{frag}} < 1$, then mass loading increases it to higher values, resulting in enhanced dust density fluctuations, further bridging the gap to the onset of GI.
    \item If the fragmentation threshold is such that $\mathrm{St_{frag}} \geq 1$, then dust densities are already high and saturated. The shift in the fragmentation barrier has no impact on dust clumping. However, the enhanced $\tilde{\mathrm{St}}_\mathrm{frag}$ from mass loading  implies less efficient radial drift and turbulent diffusion of particles, which is favorable for the onset of GI \citep{Klahr2020}.
\end{itemize}

While the initial dust size has little effect, the initial dust-to-gas ratio $\epsilon_0$ controls the efficiency of mass loading, with a higher value implying lower impact velocities between dust particles. However, if the dust-to-gas ratio at the barrier considered (bouncing or fragmentation) is too low, the SI does not increase the dust-to-gas ratio enough for the mass loading to significantly shift the barrier. 

In conclusion, for low fragmentation velocities previously believed to be prohibitive, the onset of the gravitational instability and thus the formation of planetesimals, should be facilitated by the two-way synergy between SI, dust growth and mass loading that we describe.

 \begin{acknowledgements}

We thank the referee for their insightful observations. V.V.-G. received support from the Annette Kade Graduate Student Fellowship Program of the
RGGS at the American Museum of Natural History, through generous contributions of the
Annette Kade Charitable Trust. M.-M.M.L. is partly supported by NASA grant 80NSSC25K7117. U.L. and V.V.-G. thank 
the ERC for support by synergy grant ECOGAL with grant number 855130.

\end{acknowledgements}

\bibliographystyle{aa}
\bibliography{ref}

\begin{appendix}

\section{Momentum equations}
\label{appendix:momentum equations}
We model a local patch of the protoplanetary disk centered on the midplane within the shearing box framework \citep{Goldreich1978}. The local frame co-rotates with the disk at an angular velocity $\Omega \vec{e_z}$ at an arbitrary distance $r$, where the $x$-axis is in the radial direction and the $z$-axis is vertical. The momentum conservation equations for the gas and the dust are:
 \begin{align}
  \frac{\partial \rho_\mathrm{g}  \vec{v}}{\partial t}  + \nabla \cdot \left[ \rho_\mathrm{g}  \vec{v} \vec{v} + P \mathbb{I} \right]&= -\rho_\mathrm{g}  \left[2 \vec{v} \times \Omega \vec{e_z} + 2 q \eta \Omega^2 r \vec{e_x} \right] +
  \vec{F}_\mathrm{drag} , \nonumber\\
  \frac{\partial \rho_\mathrm{d} \vec{v}_\mathrm{d}}{\partial t}  + \nabla \cdot \left[ \rho_\mathrm{d} \vec{v}_\mathrm{d} \vec{v}_\mathrm{d} \right]&= -2 \rho_\mathrm{d} \vec{v}_\mathrm{d} \times \Omega \vec{e_z}  - \vec{F}_\mathrm{drag},
  \label{eq:hydro}
\end{align}
where $P = \rho_\mathrm{g}  c_\mathrm{s}^2$ is the locally isothermal equation of state. The aspect ratio of the disk is taken constant with radius and equal to $\left(H/r\right) = 0.1$ and $q=3$. The dimensionless measure of sub-Keplerian rotation is expressed through $\eta = \Pi \left(H/r\right) = 0.05 \left(H/r\right).$

{\ttfamily SHARK} relies on the monotonic-upstream-centered scheme for conservation laws predictor-corrector scheme to solve the Riemann problems at each cell interface. The gas fluid equations are solved using an HLLC approximate Riemann solver, while the dust fluid equations are solved with the solver presented in \cite{Huang&Bai2022}. Although it is more accurate to use a common wave fan for both the dust and the gas when the two are strongly coupled, it is no longer recommended when the relative velocities are high \citep[i.e., when the decoupling scale is resolved, see][]{Verrier2025a}. Since we study here dust decoupling and clumping, we use distinct wave fans with no sound wave for the dust fluid. The drag solver is the implicit one presented in \cite{Krapp2020}. 

\section{Fragmentation velocity, domain size and grain intrinsic density}
\label{sect:domain size}

\begin{figure}
\centering
\includegraphics[width=0.5\textwidth]{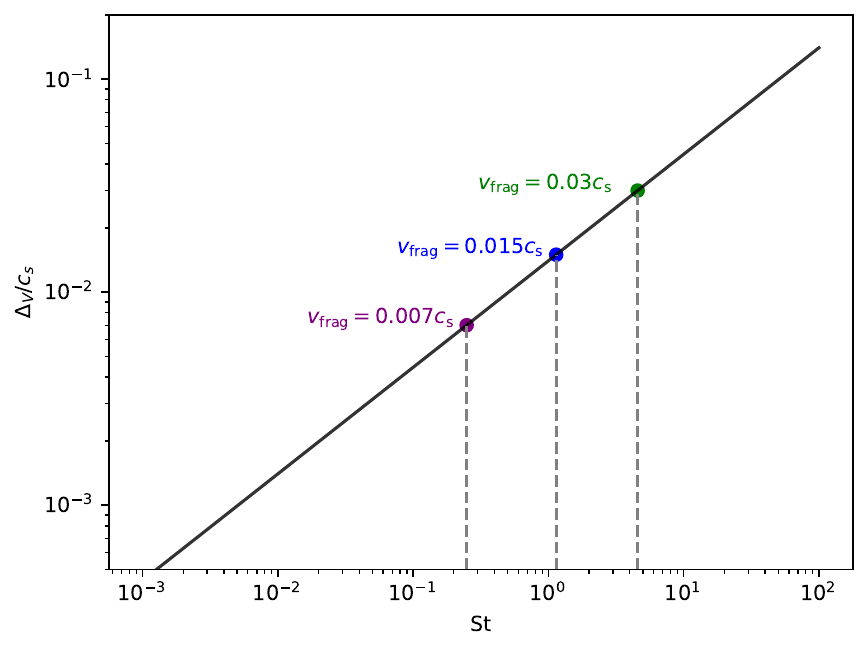}

\caption{Dust collision velocity driven by turbulence with respect to the Stokes number St of the larger grain, as predicted by the analytical model of \cite{Ormel2007}, without mass loading. The colored circles refer to the fragmentation velocity values explored in this work. The maximum St to which the dust distribution is expected to grow can be inferred. With mass loading, this limit can be increased.}
\label{fig:turb velocity vs St}
\end{figure}

In order to set realistic values of fragmentation velocities, we obtain physical estimates of the sound speed at $r = 1$ AU with a central star of mass $M = 1 \ M_\odot$. With $v_\mathrm{k} = r \Omega = (\mathcal{G} M / r)^{1/2}$ and our chosen aspect ratio of $H/r = 0.1$, we find $c_\mathrm{s} \simeq 3 \times 10^5 \ \mathrm{cm \ s^{-1}}$. A widely used value for the fragmentation threshold of bare silicate grains can be found in \cite{Ormel2009}, where $v_\mathrm{frag} = 21.2 \ \mathrm{m \ s^{-1}} = 7 \ \times 10^{-3} c_\mathrm{s}$, while for ice-coated grains $v_\mathrm{frag} = 424.3 \ \mathrm{m \ s^{-1}}  = 1.4 \ \times 10^{-1} c_\mathrm{s}$. However, recent laboratory experiments revised the dust elastic properties and found that bare silicate grains are more resilient than previously believed \citep[][]{Kimura2015,Steinpilz2019}. We thus take $v_\mathrm{frag} = 7 \ \times 10^{-3} c_\mathrm{s}$ as a lower limit and explore higher values, in the range $v_\mathrm{frag} = \left[7 \ \times 10^{-3}, 1.5 \ \times 10^{-2} \right] c_\mathrm{s}$. 

We show in Fig.~\ref{fig:turb velocity vs St} the correspondence between the dust collision velocity driven by turbulence with respect to the Stokes number St of the larger grain, as predicted by the analytical model of \cite{Ormel2007}, without mass loading. The collision velocity is given by:
\begin{equation}
\label{Ormel eq}
    \Delta v_{{i,j}} = V_\mathrm{g}\left(\beta{\left[x_{i,j}\right]}  \mathrm{St}_i\right)^{1/2},
    \end{equation}
    where $i$ and $j$ are the indices of the two colliding grains. For equal-size grains 
    \begin{equation}
         \Delta v_{{i,j}} \simeq 
    c_\mathrm{s} (1.97\alpha \mathrm{St}_i)^{1/2}. \label{Ormel approx}
\end{equation}
With mass loading, the lower effective sound speed $\tilde{c}_\mathrm{s}$ (Eq.\ \ref{modified sound speed}) is substituted in Eq.\ (\ref{Ormel approx}).  In regions where the dust-to-gas ratio $\epsilon$ is enhanced, the sound speed is effectively reduced and thus so are the turbulent collision velocities. As a consequence, higher dust sizes and St can be reached before fragmentation occurs. 
 
Regarding the domain size, we rely on \cite{Zhu&Yang2021} to estimate the wavelength of the fastest growing modes. When $\epsilon_0 \geq 1$, they find that the corresponding dimensionless wavenumber is $K_x \simeq K_z \geq 10K_{a0}$ where 
\begin{equation}K_{a} = \frac{\left(1 + \epsilon\right)^2 + \mathrm{St}^2}{2\left(1 + \epsilon\right)\mathrm{St}},
\end{equation}
as defined in \citet{Squire2018}, reaches a minimum at $\mathrm{St} = 1 + \epsilon$. $K_{a0}$ is $K_{a}$ with $\epsilon=0$. We thus use $K_x \simeq K_z = 10K_{a0}$ and set the domain size to $L_x = L_z = 3 \lambda_x = 3 (2\pi /  K_x) =3(0.031)H = 0.094H$. When $\epsilon_0 < 1$, $K_x \sim K_{a0}$ and thus $L_x = L_z = 3 \lambda_x = 0.94H$. The low numerical cost of the monodisperse approach for the treatment of dust growth allows us to work with a high resolution of $N_x=N_z=2048$ cells.

In this work, the Stokes number St of solid particles is determined directly by the dust size $s_\mathrm{d}$, and thus varies accordingly as particles grow. We make the assumption that the intrinsic density $\rho_\mathrm{gr}$ is kept fixed during the simulations, which is worth commenting on. \newline  
Indeed, the grain intrinsic density is an important parameter whose effect is twofold. First, it controls the number density $n_\mathrm{d} \propto \rho_\mathrm{gr}^{-1}$ of solid particles. For a given dust size, a higher intrinsic density implies fewer particles which in turn tends to increase the coagulation timescale (see Eq. \ref{eq:t_coag}) and thus reduce growth rates. Second, this parameter is involved in the definition of St via the stopping time $t_\mathrm{s} \propto \rho_\mathrm{gr}$. An increased intrinsic density means a higher inertia and thus a weaker coupling to gas molecules. It follows that the Stokes number St and thus the turbulent induced collision velocities should be higher (for a given dust size). However, we note that the overall effect of a higher $\rho_\mathrm{gr}$ is to increase the coagulation timescale $t_\mathrm{coag} \propto \sqrt{\rho_\mathrm{gr}}$, i.e., to slow down dust growth. Conversely, a fractal growth leading to a decreasing intrinsic density would proceed faster. \newline As already mentioned, $\mathrm{St} \propto \rho_\mathrm{gr}$ for a given dust size. As a consequence, compaction of dust grains upon collision should gradually enhance collision velocities (see Eq. \ref{Ormel eq}). However, our results should be unaffected since we work directly with the dimensionless St. The only difference is that the Stokes number should follow a different evolutionary path as both grain size and intrinsic density evolve over time, but fragmentation is expected to occur anyway at the threshold $\mathrm{St}_\mathrm{frag}$ defined by the choice of $v_\mathrm{frag}$ (which depends on the dust elastic properties and shape). In protoplanetary disks, pebbles close to the fragmentation barrier are expected to have already gone through compaction, at about $s_\mathrm{d} \sim \mathrm{mm}$ for bare-silicate grains and $s_\mathrm{d} \sim \mathrm{cm}$ for ice-coated grains \citep{Blum2018}. Nevertheless, this point should be more relevant and thus addressed for dust grain evolution prior to the bouncing barrier, where dust growth could be fractal.
\section{Validation tests}
\label{Sect:validation tests}

\begin{figure}
    \includegraphics[width=0.5\textwidth]{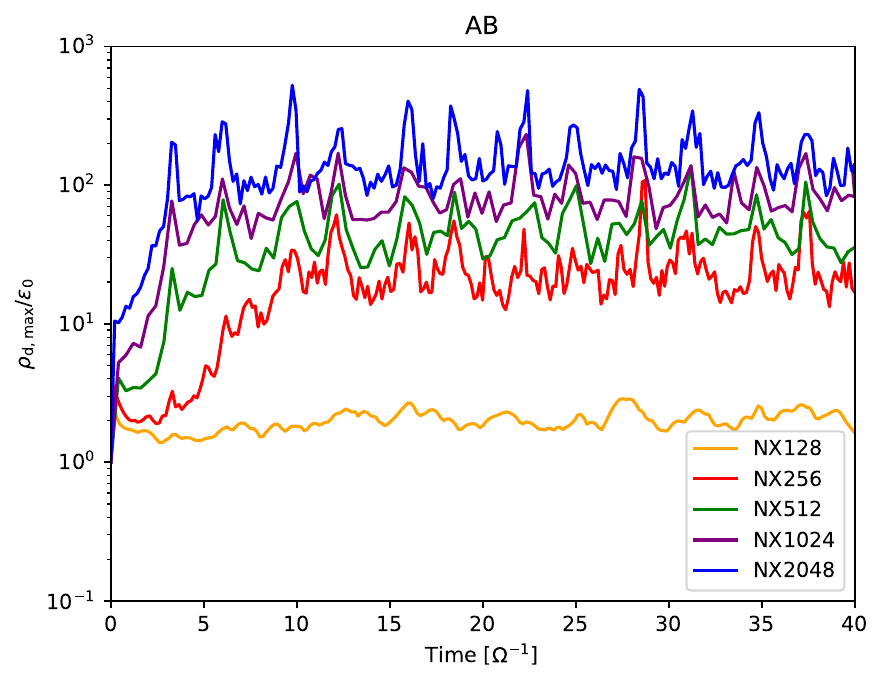}
    \caption{Maximum dust density divided by initial dust-to-gas ratio as a function of time for different spatial resolutions for the AB model defined in \cite{Johansen2007}: fixed $\mathrm{St}=0.1$ and $\epsilon_0 = 1$.}
    
\label{fig: AB test}
\end{figure}

\begin{figure}
    \includegraphics[width=0.5\textwidth]{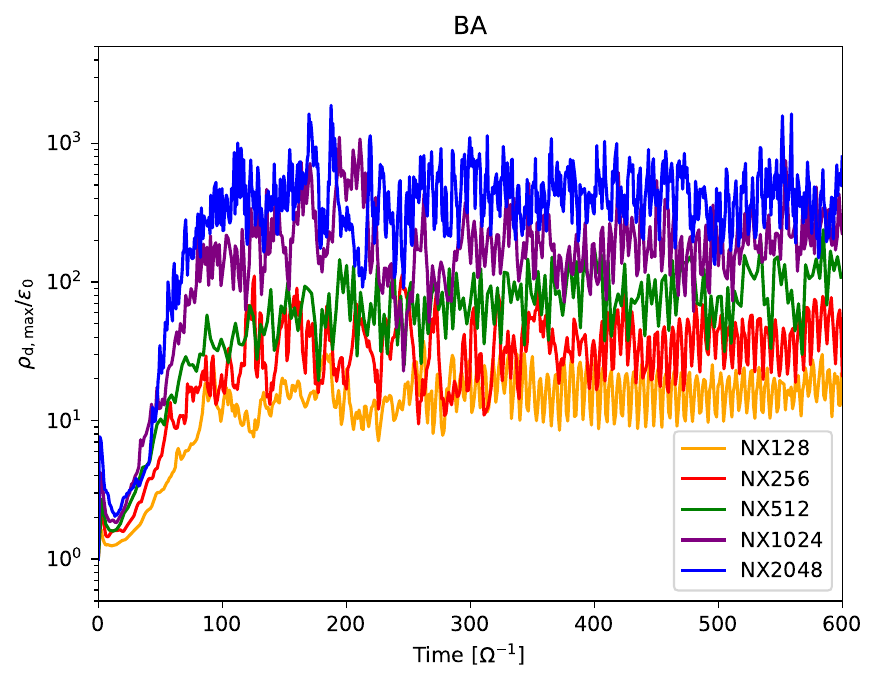}
    \caption{Same as \ref{fig: AB test} but for the BA model defined in \cite{Johansen2007}: fixed $\mathrm{St}=1$ and $\epsilon_0 = 0.2$.}
    
\label{fig: BA test}
\end{figure}

We provide a test of the code by running SI simulations with a single dust species of fixed Stokes number and specific initial dust-to-gas ratio corresponding to the models defined in \citet{Johansen2007} AB, with $\mathrm{St}=0.1$ and $\epsilon_0 = 1$ and BA, with $\mathrm{St}=1$ and $\epsilon_0 = 0.2$. As in Fig.\ 11 of \cite{LLambay2019}, we compare the maximum dust density divided by initial dust-to-gas ratio as a function of time for various resolutions. Our results agree well with theirs: Fig.\ \ref{fig: AB test} shows continued growth in the maximum value with increasing resolution for the AB test, while Fig\ \ref{fig: BA test} shows convergence at values below $10^3$ for the BA test. Ne note only slight discrepancies in terms of maximum dust density at very low resolutions,x
presumably due to the use of different drag and Riemann solvers. Again in agreement with \citet{LLambay2019}, growth rates prior to saturation also increase with resolution in the AB case, but not in the BA case for $\mathrm{N}_x \geq 512$.

\section{Impact of initial dust-to-gas ratio $\epsilon_0$}

\begin{figure}[]
    \includegraphics[width=0.5\textwidth]{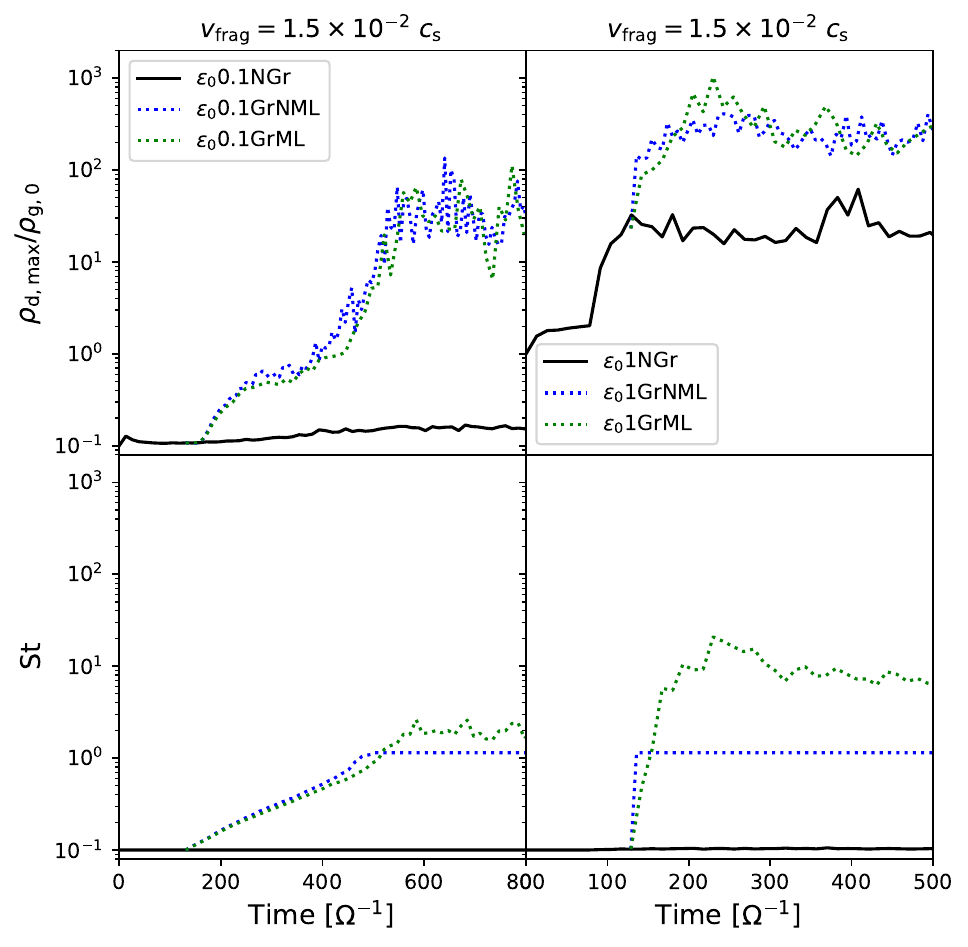}
    \caption{Evolution of the maximum dust-to-gas ratio {\em (top row)} and St {\em (bottom row)} with time for the different dust growth scenarios NGr, GrNML, and GrML. The models in the {\em left} column use an initial dust-to-gas ratio of $\epsilon_0 = 0.1$ while in the {\em right} column they use $\epsilon_0 = 1$. The velocity fragmentation threshold $v_\mathrm{frag} = 1.5\times 10^{-2} c_\mathrm{s}$ in both cases.}
    
\label{fig: rhod and St for varying eps}
\end{figure}

\begin{figure}[]
    \centering
    \includegraphics[width=0.52\textwidth]{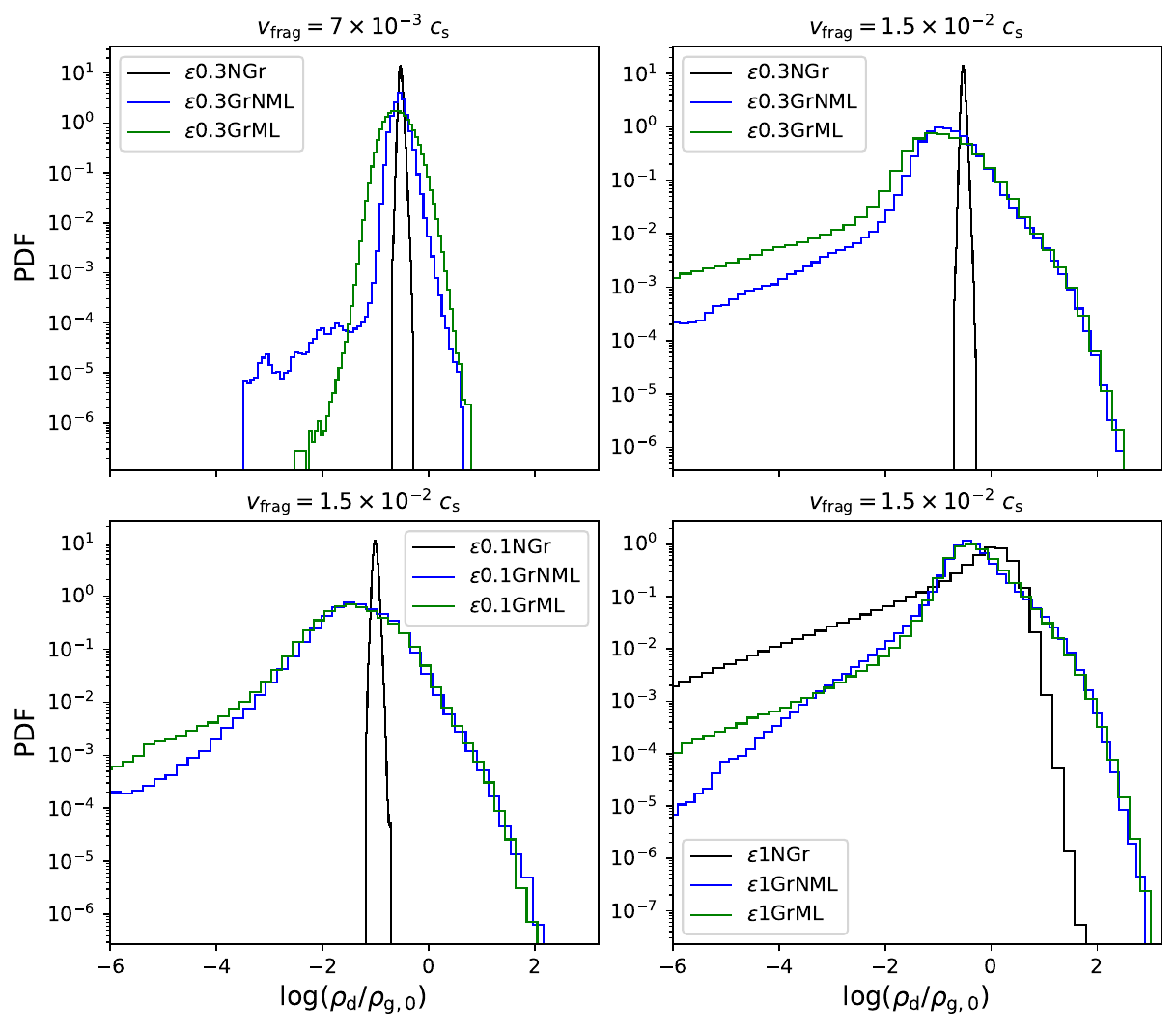}
    \caption{Time-averaged  PDF of the dust-to-gas ratio in the saturated regime of the SI for different values of initial dust-to-gas ratio $\epsilon_0$ and fragmentation velocity $v_\mathrm{frag}$. Different dust growth scenarios are explored: NGr, GrNML, and GrML. }
    
\label{fig:1D pdf}
\end{figure}

We explore here the influence of the initial dust-to-gas ratio, by comparing a lower value $\epsilon_0 = 0.1$ to a higher value $\epsilon_0 = 1$, with fixed fragmentation velocity $v_\mathrm{frag} = 1.5\times 10^{-2} c_\mathrm{s}$.  
For this value of $v_\mathrm{frag}$, we saw in Sect.\ \ref{subsec:loose} that any increase in the Stokes number beyond the limit $\mathrm{St_{frag}} \geq 1$ does not produce stronger dust density fluctuations due to efficient decoupling of the dust fluid, which is confirmed in Fig.\ \ref{fig: rhod and St for varying eps} for both values of $\epsilon_0$. The effect of changing $\epsilon_0$ is threefold. First, a higher value implies a stronger shift of the fragmentation barrier due to the dust fluid being more massive and thus significantly reducing the effective sound speed of the mixture. Indeed, the shifted barrier lies around $\tilde{\mathrm{St}}_\mathrm{frag} \simeq 2$ for $\epsilon_0 = 0.1$ but climbs to $\tilde{\mathrm{St}}_\mathrm{frag} \simeq 10$ for $\epsilon_0 = 1$. Second, dust growth rates are increased for larger $\epsilon_0$ (see Eq.\ \ref{eq:t_coag}) where we see in Fig.\ \ref{fig: rhod and St for varying eps} that the Stokes number rises by two orders of magnitude in barely 50 orbital times for $\epsilon_0 = 1$ while more than 50 orbital times are needed to reach the fragmentation barrier for $\epsilon_0 = 0.1$. Finally, the SI is known to generate stronger dust clumping as $\epsilon_0$ increases, as depicted in Fig.\ \ref{fig: rhod and St for varying eps}. 

In Fig.\ \ref{fig:1D pdf}, we present the time-averaged probability density function (PDF) of dust-to-gas ratio in the saturated regime for different sets of parameters. As already suggested by the time evolution of the maximum value in Figs.\ \ref{fig: rhod and St for eps0.3} and~\ref{fig: rhod and St for varying eps}, we see that the runs with dust growth (GrNML and GrML) produce a strikingly larger dispersion of dust-to-gas ratio. In addition, for $v_\mathrm{frag} = 7\times 10^{-3} c_\mathrm{s}$, mass loading (GrML) leads to a higher occurrence of larger values than the GrNML run. In contrast, the right-hand tail of the distributions are overlapping for both runs when $v_\mathrm{frag} = 1.5\times 10^{-2} c_\mathrm{s}$, confirming that the decoupled grains do not further enhance density fluctuations. 

\section{A glimpse at the bouncing barrier}

\begin{figure}[]
    \includegraphics[width=0.5\textwidth]{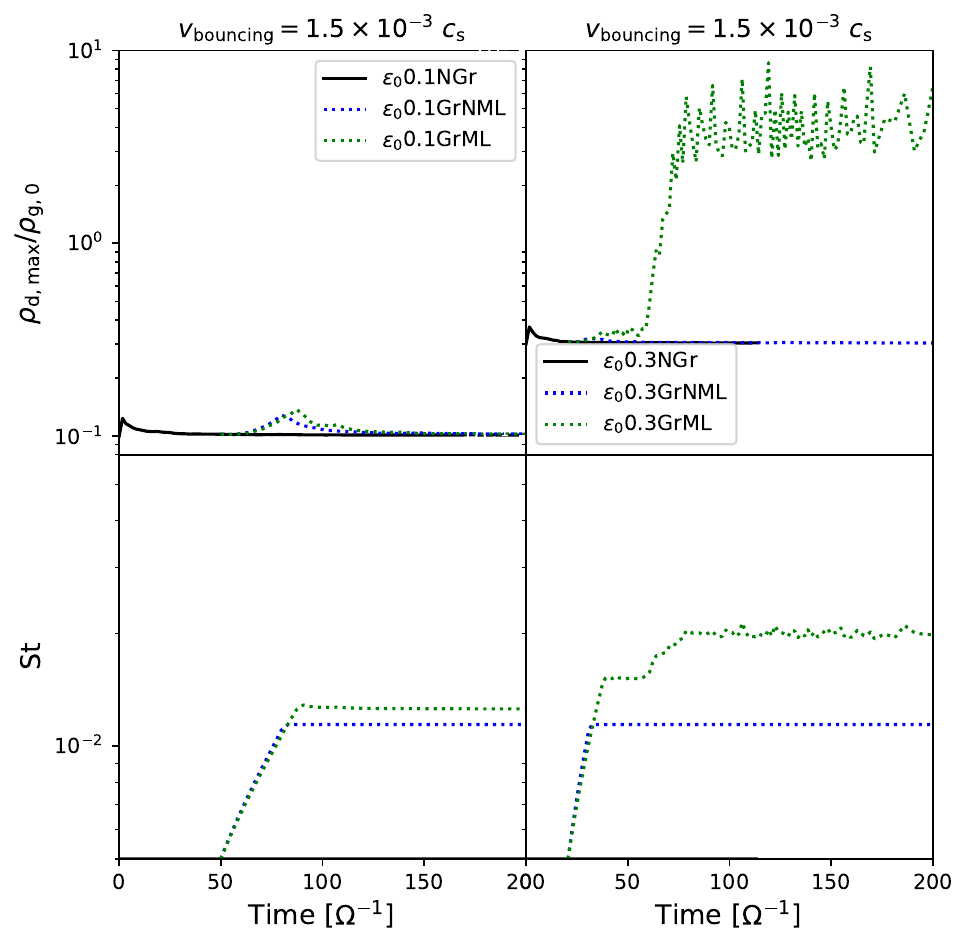}
    \caption{Same as Fig.\ \ref{fig: rhod and St for varying eps} but starting from a lower initial Stokes number $\mathrm{St_{ini}} = 10^{-2}$ with a bouncing barrier set at the value $v_\mathrm{frag} = 1.5\times 10^{-3} c_\mathrm{s}$. The left column corresponds to an initial dust-to-gas ratio of $\epsilon_0 = 0.1$ whereas the right column corresponds to $\epsilon_0 = 0.3$.}
    
\label{fig: rhod and St for bouncing barrier}
\end{figure}

This particular case explores the conditions associated with the bouncing barrier by considering  $\mathrm{St_{ini} = 10^{-2}}$ and a bouncing velocity $v_\mathrm{bouncing} = 1.5 \times 10^{-3} \ c_\mathrm{s}$ with a value ten times lower than the fiducial $v_\mathrm{frag}$. With the lower value $\epsilon_0 = 0.1$, we see that almost no dust clumping is produced even with mass loading, due to the maximum St remaining low. However, a larger value of $\epsilon_0 = 0.3$  allows the dust density to soar in the mass loading (GrML) run, reaching a value close to $\epsilon 
\simeq 5$ as the effective Stokes number grows to $\tilde{\mathrm{St}}_\mathrm{frag} \simeq 2 \times 10^{-2}$. This situation clearly highlights the important role of the initial dust-to-gas ratio $\epsilon_0$.

\section{How do the Stokes number and the dust-to-gas ratio correlate}
\label{sect:2D pdf}

\begin{figure*}[hbt]
    \centering
    \includegraphics[width=0.8\textwidth]{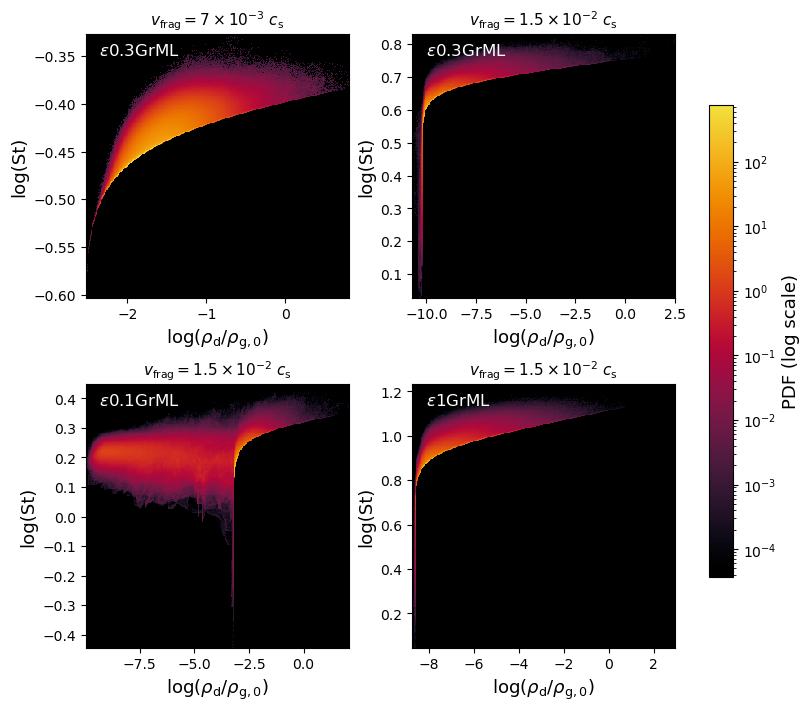}
    \caption{Bivariate PDF of the Stokes number St of the dust fluid and the dust-to-gas ratio $\rho_\mathrm{d} / \rho_\mathrm{g,0}$ for different values of initial dust-to-gas ratio $\epsilon_0$ and fragmentation velocity $v_\mathrm{frag}$.}
\label{fig:2D pdf}

\end{figure*}

Figure \ref{fig:2D pdf} shows the bivariate PDF of the Stokes number St of the dust fluid and the dust-to-gas ratio $\rho_\mathrm{d} / \rho_\mathrm{g,0}$ for different combinations of initial dust-to-gas ratio $\epsilon_0$ and fragmentation velocity $v_\mathrm{frag}$, with dust growth and dust mass loading (GrML).

As expected, there is a clear correlation between both quantities for $v_\mathrm{frag} = 7 \ \times 10^{-3} c_\mathrm{s}$, when $\tilde{\mathrm{St}}_\mathrm{frag} < 1$, since in this loosely coupled regime, higher values of St yield stronger dust density fluctuations (see Fig.\ \ref{fig: rhod and St for eps0.3}). However, for the higher fragmentation velocity $v_\mathrm{frag} = 1.5 \ \times 10^{-2} c_\mathrm{s}$, the fragmentation barrier is reached at $\tilde{\mathrm{St}}_\mathrm{frag} > 1$ and the coupling between dust growth and clumping is only one way because particles with St above unity decouple from the gas. Overdensities promote dust growth, but in turn, the larger dust grains formed do not produce stronger density fluctuations via the SI. This is reflected in the PDF as we see that for $v_\mathrm{frag} = 1.5 \ \times 10^{-2} c_\mathrm{s}$, the counts are more uniform for most of the combination of values, and the more likely values show that a large range of dust-to-gas ratio can be associated with the same Stokes number. 
    
\end{appendix}

\end{document}